\documentclass[notitlepage,a4paper,aps,prd,onecolumn,superscriptaddress,nofootinbib,groupedaddress,longbibliography]{revtex4-1}
\usepackage{amsmath}
\usepackage{amsfonts}
\usepackage{amssymb}
\usepackage[utf8]{inputenc}
\usepackage[T1]{fontenc}
\usepackage[colorlinks=true]{hyperref}
\usepackage[dvipsnames]{xcolor}
\usepackage{graphicx}
\usepackage[normalem]{ulem}
\usepackage{enumerate}

\renewcommand{\d}{\mathrm{d}}

\usepackage[colorlinks=true]{hyperref}
\newtheorem{defi}{Definition}

\begin{document}
	
\title{Finsler spacetime geometry in Physics}

\author{Christian Pfeifer}
\email{christian.pfeifer@ut.ee}
\affiliation{Laboratory of Theoretical Physics, Institute of Physics, University of Tartu, W. Ostwaldi 1, 50411 Tartu, Estonia}

\begin{abstract}
Finsler geometry naturally appears in the description of various physical systems. In this review I divide the emergence of Finsler geometry in physics into three categories: as dual description of dispersion relations, as most general geometric clock and as geometry being compatible with the relevant Ehlers-Pirani-Schild axioms. As Finsler geometry is a straightforward generalisation of Riemannian geometry there are many attempts to use it as generalized geometry of spacetime in physics. However, this generalisation is subtle due to the existence of non-trivial null directions. I review how a pseudo-Finsler spacetime geometry can be defined such that it provides a precise notion of causal curves, observers and their measurements as well as a gravitational field equation determining the Finslerian spacetime geometry dynamically. The construction of such Finsler spacetimes lays they foundation for comparing their predictions with observations, in astrophysics as well as in laboratory experiments.
\end{abstract}

\maketitle


\section{Introduction}\label{sec:intro}
Finsler geometry is a straightforward generalization of Riemannian geometry. Instead of deriving the geometry of a manifold from a Riemannian metric $g$, its Levi-Civita connection and the corresponding induced length measure $F_g(X)=\sqrt{g(X,X)}$ for vectors $X$, the geometry is derived from a general $1$-homogeneous length measure $F(X)$, called the Finsler function, and its Cartan non-linear connection. The first idea of going beyond the quadratic length measure was already mentioned by Riemann himself in his inaugural lecture \cite{Riemann}, but only Finsler started to investigate the properties of spaces on which the length of curves is measured by a general $1$-homogeneous length measure systematically \cite{Finsler}. Since then, Finsler geometry became an established subject in mathematics \cite{Bao, Miron}.

To apply Finsler geometry as geometry of spacetime in physics, i.e.\ as extension of pseudo-Riemannian geometry with a metric of Lorentzian signature, is not as straightforward as it is in the case of Riemannian geometry. The existence of non-trivial vectors of zero Finsler length causes non-smoothness and degeneracies of the Finsler geometry, which have to be investigated carefully. Physically the existence of such null-vectors is however necessary, since they are the tangent vectors to curves which describe the propagation of massless particles, e.g.\ light.

The aim of this article is to summarize how a Finslerian spacetime geometry can be constructed and how to obtain a consistent theory of gravity based on Finsler spacetime geometry.

To understand what a Finslerian spacetime geometry must necessarily provide for physics lets recall that the geometry of spacetime plays a three-fold role:
\begin{itemize}
	\item it defines a causal structure of events, i.e.\ it classifies which events on spacetime have been able to influence a particular event, and which events in the future can be influenced by a particular event;
	\item it defines observers and their measurements;
	\item it encodes the gravitational interaction and its dynamics.
\end{itemize}
In general relativity this three-fold role of the geometry of spacetime is realized by equipping the spacetime manifold $M$ with a metric tensor $g$ of Lorentzian signature, where:
\begin{itemize}
	\item the causal structure is defined by the causal (timelike and lightlike) curves, identified by the Lorentzian metric;
	\item an observer is defined as proper time parametrized timelike curve $\gamma$ on $M$, equipped with a frame $\{e_\mu \}_{\mu=0}^3$ of the metric $g$. The unit time direction of the observer is given by $e_0 = \dot \gamma$ and the $\{e_\alpha\}_{\alpha=1}^3$ define the observer's spatial unit length directions;
	\item gravity and its dynamics are described by the Einstein equations $r_{ab}-\frac{1}{2}g_{ab}r = \frac{8\pi G}{c^4}t_{ab}$ and the minimal coupling between matter fields and spacetime geometry. Here $r_{ab}$ and $r$ are the Ricci tensor and Ricci scalar of the Levi-Civita connection of the metric, and $t_{ab}$ is the energy momentum tensor of the matter fields on spacetime.
\end{itemize}
Whenever one starts to modify the geometry of spacetime away from the Lorentzian metric one employed in general relativity, for example in the search for geometric explanations of dark energy and dark matter or to  approach a theory of quantum gravity, one has to keep in mind that changing the geometry of spacetime may effect all of its three roles. Hence, in order to compare a modified spacetime geometry with observations, it is necessary to investigate how the causal structure, observers and their measurements as well as gravity and its dynamics are obtained from the modified geometry of spacetime consistently. How this can be done for a Finslerian spacetime geometry is subject of this article, which is structured as follows:

I will begin by recapitulating instances where and how Finsler spacetime geometry naturally emerges in the description of physical systems in Section \ref{sec:fip}. Afterwards I recall definitions of Finsler spacetimes and focus on a most recent one, which combines the most prominent existing ones. I discuss how this new definition realizes a well defined causal structure in Section \ref{sec:causal}, before I define observers and their measurement on Finsler spacetimes in Section \ref{sec:obs} as well as recall results on the gravitational dynamics for Finsler spacetimes in Section \ref{sec:gravi}. In the end, in Section~\ref{sec:concl}, I will conclude that it is possible to realize the three fold role of the geometry of spacetime with Finsler geometry and list open problems and next steps in the program of Finsler spacetime geometry.

Throughout this article the following notation will be used. An element $Y$ of the tangent space $T_xM$ at $x\in M$ of a manifold $M$ can be expanded in a coordinate chart $(U,x)$, $U\subset M$ open, as $Y=\dot x^a\partial_a$. The union of all tangent spaces of $M$ is called the tangent bundle $T^*M=\bigcup_{x\in M}T_xM$. The tuple $(x,\dot x)$ is the coordinate representation of $Y$ in a manifold induced coordinate chart $(TU, (x,\dot x))$ on $TM$. The corresponding basis of the (co)-tangent spaces to $TM$ will be denoted by $(\partial_a=\frac{\partial}{\partial x^a}, \dot{\partial}_a = \frac{\partial}{\partial \dot x^a})$ and $(\d x^a, \d \dot x^a)$. Similarly an element of the cotangent bundle $T^*M$ of a manifold $M$ can be expanded in a coordinate chart $(U,x)$, $U\subset M$ open, as $P=p_a \d x^a$. The tuple $(x,p)$ is the coordinate representation of $P$ in a manifold induced coordinate chart $(T^*U, (x,p))$ on $T^*M$. The corresponding basis of the tangent spaces to $T^*M$ will be denoted by $(\partial_a=\frac{\partial}{\partial x^a}, \bar{\partial}^a = \frac{\partial}{\partial p_a})$.

\section{Finsler geometry in physics}\label{sec:fip}
There are mainly three lines of arguments leading naturally to a Finslerian spacetime geometry in physics: the dual description of dispersion relations, the most general mathematical realization of the clock postulate of special relativity and a modification of the axiomatic approach to spacetime geometry by Ehlers, Pirani and Schild.

All approaches demonstrate that a Finslerian geometry emerges naturally in physics. The open question is which Finslerian length measure describes nature in which situation. Is the only consistent one defined by a Lorentzian metric alone, as it is employed in special and general relativity, or can more general ones be viable? (See also the discussion in \cite{Goenner:2008rr,Lammerzahl:2018lhw}.)

\subsection{Duals of dispersion relations}
A dispersion relation is a constraint the four-momentum of a physical particle has to satisfy. It is described in terms of a Hamilton function on the point particle phase space of spacetime. The dual description in terms of the particle's velocity instead of its momentum gives rise to a Finsler Lagrangian.

Mathematically, dispersion relations are level sets of a Hamilton function $H$ on the cotangent bundle $T^*M$ of a, usually four dimensional, spacetime manifold $M$,
\begin{align}
	H: T^*M \rightarrow \mathbb{R};\quad (x,p) \mapsto H(x,p)\,,\quad H(x,p) = \epsilon\,,
\end{align}
where $\epsilon$ is a constant. Physical point particles propagate on trajectories $(x(\tau),p(\tau))$ in $T^*M$, which are solutions of Hamilton's equations of motion
\begin{align}\label{eq:hameom}
	\dot x^a = \bar\partial^aH, \quad \dot p_a = - \partial_a H\,.
\end{align}
Trajectories satisfying $H(x,p) = 0$ are called massless, or lightlike The set of trajectories for which $H(x,p)\neq 0$ contains the worldlines of massive particles. However, in general, additional conditions must be posed to identify these physical trajectories. The precise form of these extra conditions depends on the class of dispersion relations in consideration. 

Prominent examples of physically interesting Hamilton functions in the literature are:
\begin{itemize}
	\item General bi-hyperbolic polynomial dispersion relations based on a $(n,0)$-tensor field on spacetime, emerging from well-defined quantizable field theories, see \cite{Raetzel:2010je},
	\begin{align}
	H_{poly}(x,p) = G^{a_1...a_n}(x)p_{a_1}...p_{a_n}\,.
	\end{align}
	A particular interesting special case is the $4$th order Fresnel polynomial derived from local and linear premetric electrodynamics, which describes the propagation of light in local and linear premetric electrodynamics \cite{Hehl}
	\begin{align}
	H_{Fresnel}(x,p) = \mathcal{G}^{a_1...a_4}(x)p_{a_1}...p_{a_4}\,,
	\end{align}
	where $\mathcal{G}^{a_1...a_4}(x)$ is the Fresnel tensor introduced in \cite{Rubilar:2007qm}. A specific type of dispersion relations, appearing in this context, are for example bi-metric ones. They describe the propagation of light inside birefringent crystals \cite{Perlick}. As the name suggests, they are built from a product of two Lorentzian metrics $g$ and $h$, whose inverses have components $g^{ab}(x)$ and~$h^{ab}(x)$ 
	\begin{align}
	H_{bi}(x,p) = (g^{ab}(x)p_ap_b)(h^{cd}(x)p_cp_d)\,.
	\end{align}
	\item Non-homogeneous, quantum gravity phenomenology inspired dispersion relations \cite{Mattingly:2005re,AmelinoCamelia:2008qg,Liberati:2013xla}, like the $\kappa$-Poincar\'e dispersion relations \cite{Lukierski:1991pn}, displayed here on curved spacetime, see \cite{Barcaroli:2017gvg},
	\begin{align}
	H_\kappa(x,p) = -\tfrac{4}{\ell^2}\sinh(\tfrac{\ell}{2}Z^a(x)p_a)^2 + e^{\ell Z^a(x)p_a}(g^{ab}(x)p_ap_b + (Z^a(x)p_a)^2)\,,
	\end{align}
	where $g^{ab}(x)$ and $Z^a(x)$ are the components of a Lorentzian metric and a unit timelike vector field with respect to $g$. The parameter $\ell$ is usually identified with the Planck length.
	\item Non-relativistic spatial dispersion relations for waves in media, for example used in the description of seismic waves \cite{Yajima2009,Klimes,Cerveny}, written in a four dimensional space-time language
	\begin{align}
	H(x,p) = (Z^a(x) p_a)^2 - \bar H(x,\vec p)\,, 
	\end{align}
	where again $Z^a(x)$ are the components of vector field that is interpreted as time orientation and $\bar H(x,\vec p)$ is a spatial Hamiltonian which only depends on the momenta $\vec p(p)$ which are co-normal to $Z$, i.e. $\vec p_aZ^a(x) = 0$. This means for a given four momentum with components $p_a$ we have $\vec p_a(p) = p_a - p_bZ^b(x) \kappa_a(x)$, where $\kappa_a(x)$ are the components of a dual $1$-form field to $Z$, i.e.\ $Z^a(x)\kappa_a(x)=1$.
\end{itemize}

Wherever the relation between velocities and momenta (the first equation in \eqref{eq:hameom}) is invertible\footnote{Locally, the requirement is that the Hessian of the Hamilton function $H$ with respect to the momenta is non-degenerate}, a dispersion relation, which describes the propagation of massive and massless point particles in terms of their positions and momenta on the cotangent bundle, can be mapped to a description of the particles' motion in terms of their positions and velocities on the tangent bundle. To do so we employ the Helmholtz action
\begin{align}
	S_H[x,p,\lambda] = \int \d \tau\ (\dot x^a p_a - \lambda\ f(H(x,p),\epsilon) )\,,
\end{align}
where $\lambda$ is a Lagrange multiplier which ensures that particles satisfy the dispersion relation, $f(H,\epsilon)$ is a function such that $f(H(x,p),\epsilon) = 0$ implies $H(x,p) = \epsilon$ and $\epsilon$ characterizes if massless or massive particles are considered.

Variation of the action with respect to $\lambda$ yields the dispersion relation, variation with respect to $p$ allows us to solve the corresponding equation of motion for $p(x,\dot x,\lambda)$ and to obtain a new action $\bar S[x,\lambda] := S[x,p(x,\dot x, \lambda),\lambda]$. Variation of $\bar S$ with respect to $\lambda$ and solving for $\lambda$ as function of $x$ and $\dot x$ finally yields the point particle action
\begin{align}
	S[x] := \bar S[x,\lambda(x,\dot x)] = \int \d \tau\ F_\epsilon(x,\dot x)\,.
\end{align}
For massive particles, i.e.\ $\epsilon\neq 0$, $F_\epsilon$ is a $1$-homogeneous Finsler function, for massless particles $F_0$ must not necessarily be $1$-homogeneous.

The transition from a dispersion relation to a Finslerian geometry has been worked out explicitly for Planck scale modified dispersion relations \cite{Amelino-Camelia:2014rga,Letizia:2016lew} and for weakly premetric electrodynamics \cite{Gurlebeck:2018nme}. Depending on the Hamiltonian from which one starts a huge variety of Finsler functions can be obtained.
 
\subsection{The most general geometric clock}
A cornerstone of special relativity are the two axioms:
\begin{itemize}
	\item The constancy of the speed of light, i.e.\ the speed of light has the same constant value for any inertial observer, independent of its relative motion with respect to the light source.
	\item The clock postulate, i.e.\ the time which passes for an observer between two events is given by the length of its worldline connecting these two events.
\end{itemize}
The first axiom must be realized by a precise description of how observers measure velocities and hence the speed light, while the second axiom, on which we focus here, leads directly to a Finslerian spacetime geometry.

In order to to realize the clock postulate mathematically it is necessary to equip the spacetime manifold with a length measure for observer worldlines $x(\tau)$. The most general geometric length measure $S$ is defined in terms of a general Finsler function $F$
\begin{align}\label{eq:Finsler}
	S[x] = \int \d \tau\ F(x,\dot x)\,.
\end{align}
Technically $F$ is a function on the tangent bundle of $M$ satisfying $F(x,\lambda\dot x) = \lambda F(x,\dot x)\ \forall \lambda >0$, so that the length measure is parametrization invariant. The question to be answered is, which $F$ is physically viable? 

One principle, which may be applied to identify a physical length measure, is local symmetries and observer transformations. Demanding that the length measure is locally Lorentz invariant yields that it must be defined in terms of a Lorentzian metric $g$ as
\begin{align}
	F = \sqrt{|g_{ab}(x)\dot x^a \dot x^b|}\,.
\end{align}
Following the reasoning towards special and general relativity historically, the emergence of local Lorentz invariance can nicely be traced back to the combination of Maxwell-Lorentz electrodynamics in vacuum and the first axiom of special relativity~\cite{ThePrincipalOfRelativity}. 

Relaxing this symmetry demand by considering a length measure which is invariant under all transformation which leave the massless metric wave equation invariant one reaches the Bogoslovsky \cite{Bogoslovsky1977}, or very special/general relativity~\cite{Cohen:2006ky,Gibbons:2007iu}, length measure, defined in terms of a Lorentzian metric $g$ and a $1$-form $A$,
\begin{align}\label{eq:bogos}
	F = (|g_{ab}(x)\dot x^a \dot x^b|)^{\frac{1-q}{2}} (A_c(\dot x)\dot x^c)^q\,.
\end{align}

Instead of arguing with symmetry further length measures of interest are identified via their appearance in the description of physical systems. 

The Randers length, originally suggested as possible unified description of gravity and electromagnetism \cite{Randers}, describes the motion of a charged particle subject to an electromagnetic potential, the Zermelo navigation problem and the influence of wind on physical systems \cite{Gibbons:2011ib,MARKVORSEN2016208,Javaloyes:2018lex}. It is given in terms of a Lorentzian metric $g$ and a $1$-form $A$,
\begin{align}\label{eq:randers}
	F = \sqrt{|g_{ab}(x)\dot x^a \dot x^b|} + A_c(x)\dot x^c\,.
\end{align}

General $n$th roots of $n$th order polynomial length measures describe the propagation of massless modes of fields subject to general polynomial dispersion relations \cite{Raetzel:2010je}. The corresponding length measure is defined in terms of a $(0,n)$-tensor field $G$,
\begin{align}\label{eq:nthroot}
	F = |G_{a_1...a_n}(x)\dot x^{a_1}...\dot x^{a_n}|^{\tfrac{1}{n}}\,.
\end{align}
The $4$th order case describes the propagation of light derived from premetric instead of Maxwell-Lorentz electrodynamics \cite{Punzi:2007di},
\begin{align}
	F = |G_{a_1...a_4}(x)\dot x^{a_1}...\dot x^{a_4}|^{\tfrac{1}{4}}\,.
\end{align}
Using this latter length measure to realize the clock postulate, can be seen as a generalisation of the historical reasoning deducing the geometric structure of spacetime from Maxwell-Lorentz electrodynamics.

Thus, the clock postulate naturally leads to a Finslerian length measure. As in the case of the derivation of Finsler length elements from dispersion relations, there exists a large variety of Finsler functions which are of interest.

\subsection{From Ehlers-Pirani-Schild to Finsler geometry}
Ehlers, Pirani and Schild (EPS) deduce the pseudo-Riemannian Lorentzian metric geometry of spacetime, from a set of physical axioms they impose on a spacetime geometry, in four steps \cite{Ehlers2012}:

\begin{itemize}
	\item The first step is to deduce that spacetime is a differentiable manifold. This done by demanding the existence of smooth trajectories of massive and massless particles as well as smooth radar echoes between massive particle trajectories.
	\item The second step is to infer on the existence of a certain conformal structure, by demanding that the projective space of all directions at every point on spacetime shall decay into two connected components, when the directions of massless trajectories are removed. Moreover it is demanded that in a sufficiently small neighbourhood $V$ of a massive particle worldline, for every $p \in V$, which does itself not lie on the particle worldline, the map which maps $p$ to the product between the emission time $t_e$ and the return time $t_a$ of radar echo between the the particle worldline and $p$, $p\mapsto t_e t_r$, is at least twice differentiable. 
	\item The existence of a projective structure on spacetime is ensured by the third step, namely, demanding that through each point on spacetime, and for each timelike direction, there exist one unique massive trajectory passing through that point. Additionally each of these trajectories shall possess a parameter representation such that, in local coordinates at the point under consideration, $\ddot x = 0$ holds. 
	\item Last but not least, in the fourth step compatibility between the conformal structure deduced from the massless trajectories and the projective structure deduced from the massive trajectories is demanded, which finally yields a Lorentzian metric spacetime structure.
\end{itemize}
It has been demonstrated that there exists Finsler geometries which satisfy all of the EPS axioms except one: the twice differentiability assumption on the map $p\mapsto t_e t_r$. The following class of examples has been presented explicitly in \cite{TAVAKOL198523,Tavakol1986,Tavakol2009}
\begin{align}\label{eq:expg}
	F = e^{2\sigma(x,\dot x)}\sqrt{|g_{ab}(x)\dot x^a \dot x^b|}\,,
\end{align}
where $\sigma(x,y)$ is a non-singular $0$-homogeneous function in $\dot x$ satisfying
\begin{align}
	\partial_a \sigma - \dot x^b\Gamma^c{}_{ab}(x)\dot \partial_c\sigma = 0\,,
\end{align}
and $\Gamma^c{}_{ab}(x)$ is the Christoffel symbol derived from the metric components $g_{ab}(x)$. This condition ensures that the Finsler geometry derived from $F$ is of so called  Berwald type, i.e.\ close to a pseudo-Riemannian spacetime geometry but not trivial \cite{Szilasi2011}\footnote{Their defining property is that the canonical Cartan non-linear connection is a linear affine connection}. On spacetimes with such a Finsler length measure the worldlines of freely falling massless and massive particles are given by the arclength parametrized curves that satisfy the geodesic equation
\begin{align}
	\ddot x^a + \Gamma^a{}_{bc}(x)\dot x^b \dot x^c = 0\,.
\end{align}
These are, up to a different parametrization in the case of massive particles, the same curves as on the pseudo-Riemannian spacetime $(M,g)$.

Berwald spacetimes have been considered in context of very special/general relativity \cite{Fuster:2018djw}, the standard model extension~\cite{Edwards:2018lsn,Schreck:2015seb} and in the context of the realisation of the equivalence principle \cite{Gallego2017}. A general class of Berwald Finsler geometries is given by Finsler length measures which are defined in terms of the variables $\mathcal{A}=g_{ab}(x)\dot x^a \dot x^b$ and $\mathcal{B}=A_{a}(x)\dot x^a$, for $1$-forms $A = A_a(x) \d x^a$ that are covariantly constant with respect to the Levi-Civita connection of the metric $g$.

Hence, approaching Finsler spacetimes from the EPS axiomatic reveals at least one class of Finslerian geometries as candidates describing the geometry of spacetime, namely those which are of Berwald type and only have one light cone. If this class is the only class of Finsler spacetimes compatible with the EPS axioms is under ongoing investigation.

Moreover the physical necessity of each EPS axiom should be reconsidered in detail. For example the demand that the set of all directions shall decay into two connected components, when the directions of massless particle trajectories are removed, is known to be violated for the Finsler lengths derived from premetric local and linear electrodynamics~\cite{Lammerzahl:2004ww,Gurlebeck:2018nme}. The existence of more than two connected components of non-massless directions is connected to the emergence of birefringence. Spacetime geometry induced birefringence must not necessarily be excluded by an axiom, but can be constraint by observation. If one allows for Finsler structures with more than one light cone, depends strongly, on which of the many definitions of Finsler spacetimes in the literature one employs. The one we will present in the next section allows that the set of all directions decays into more than two connected component, when the directions of massless particle trajectories are removed.

\section{Finsler spacetimes and their causal structure}\label{sec:causal}
Throughout the huge variety of Finsler length measures which we encountered in the previous section, we seek to identify the properties they must share in order to define a viable spacetime geometry, i.e. so that they provide the three fold role of the geometry of spacetime discussed in the introduction. The definition of Finsler spacetimes lays the foundation for this task.

A general definition of a Finsler spacetime will be given as a list of properties a Finsler Lagrangian $L: TM \mapsto \mathbb{R}$ and its corresponding $L$ metric (or Finsler metric), defined by the components
\begin{align}\label{eq:gL}
	g^L_{ab} = \frac{1}{2}\dot{\partial}_a\dot{\partial}_b L\,,
\end{align}
have to satisfy, in order to establish a clear notion of causal, i.e.\ timelike and lightlike, curves. The $1$-homogeneous Finsler function $F$, usually encountered in the literature as fundamental object, will be derived from $L$ as~$F = \sqrt{|L|}$.

The first systematic definition of a Finsler spacetime as a Finslerian extension of pseudo-Riemannian, Lorentzian, geometry goes back to Beem \cite{Beem}, who demanded that the signature of the $g^L$ is Lorentzian on all of $TM\setminus{0}$. This however turned out to be too restrictive and excluded many of the examples we encountered previously. Since then, there have been numerous attempts to refine Beem's definition, in order to find a suitable definition which meets the requirements from physics. However most of the existing definitions so far turned out to be still too restrictive, since different communities formulated them having in mind different examples they wanted to cover. For example the definition by the author and collaborator in \cite{Pfeifer:2011tk} excludes Randers type metrics, since there exists no power of a Randers Finsler Lagrangian that is smooth on $TM\setminus{\{0\}}$, while $n$th root Finsler spacetimes \eqref{eq:nthroot} are excluded by the definition of L\"ammerzahl, Perlick and Hasse \cite{Lammerzahl:2012kw} and by Minguzzi \cite{Minguzzi:2014aua}, since the signature of the $L$-metric changes on $TM\setminus{\{0\}}$, as well as by the definition of Javaloyes and S\'{a}nchez \cite{Javaloyes:2018lex}, since their square is not smooth where they vanish.

The following definition of Finsler spacetimes, formulated in \cite{Hohmann:2018rpp}, summarizes conditions on $L$ and $g^L$ such that freely falling causal curves exist and the geometry is well defined along nearly all of these causal curves. It is distilled from three existing definitions, see  \cite{Pfeifer:2011tk,Lammerzahl:2012kw,Javaloyes:2018lex}, in such a way that huge classes of Finsler spacetimes according to the earlier definitions are Finsler spacetimes according to this latest definition, even if not all. Moreover, it includes most examples encountered in Section \ref{sec:fip}.

\begin{defi}[Finsler spacetimes]\label{def:fst}
By a \textit{Finsler spacetime}, we understand a pair $(M,L)$, where $L:TM\rightarrow \mathbb{R}$ is a continuous function, called the Finsler-Lagrange function, which satisfies:
\begin{itemize}
	\item $L$ is positively homogeneous of degree two with respect to $\dot x$: $L(x,\lambda \dot x) = \lambda^2 L(x,\dot x)$;
	\item $L$ is smooth and the vertical Hessian of $L$ (called $L$-metric $g^L$)
	\begin{align}\label{g_def}
	g^L_{ab} = \dfrac{1}{2} \dot{\partial}_a  \dot{\partial}_b L
	\end{align}
	is non-degenerate, on a conic subbundle $\mathcal{A}$ of $TM$ such that  $TM\setminus \mathcal{A}$ is of measure zero;
	\item there exists a connected component $\mathcal{T}$ of the preimage $L^{-1}((0,\infty))\subset TM$, such that on $\mathcal{T}$ the $L$-metric $g^L$ exists, is smooth and has Lorentzian signature $(+,-,-,-)~$\footnote{It is possible to equivalently formulate this property with opposite sign of $L$ and metric $g^L$ of signature $(-,+,+,+)$. We fixed the signature and sign of $L$ here to simplify the discussion.}
	\item the Euler-Lagrange equations
	\begin{align}\label{eq:EL}
	\frac{d}{d \tau} \dot{\partial}_a L - \partial_a L = 0\,.
	\end{align}
	have a unique local solution for every initial condition $(x,\dot x)\in \mathcal{T}\cup \mathcal{N}$, where $\mathcal{N}$ is the kernel of $L$. At points of $\mathcal{N}$ where the $L$-metric degenerates the solution must be constructed by continuous extension. This means that the geodesic equation coefficients admit a $\mathcal{C}^{1}$ extension at those points.
\end{itemize}
\end{defi}
Recall that a conic subbundle $\mathcal{Q}$ of $TM$ satisfies $\pi(\mathcal{Q}) = M$, admits local trivializations and, in the local manifold induced coordinates inherited from $TM$, $(x,\dot x)\in Q \Leftrightarrow (x,\lambda \dot x)\in Q\ \forall\ \lambda>0$ holds. The projection $\pi$ is the projection of the tangent bundle.

The difficulty in writing down a precise definition of a Finsler spacetime lies in the characterization where the Finsler Lagrangian in consideration is differentiable, where $g^L$ has which signature (in particular where it may or may not be non-degenerate) and to ensure the existence of a connected component of directions, which form a convex cone. This is characterised by four conic subbundles
\begin{itemize}
	\item $\mathcal{A}$: the subbundle where $L$ is smooth and $g^L$ is non-degenerate, with fiber $\mathcal{A}_{x} = \mathcal{A} \cap T_xM$, called the set of \emph{admissible vectors},
	\item $\mathcal{N}$: the subbundle where $L$ is zero, with fiber $\mathcal{N}_x = \mathcal{N} \cap T_xM$,
	\item $\mathcal{A}_0 = \mathcal{A}\setminus\mathcal{N}$: the subbundle where $L$ can be used for normalization, with fiber  $\mathcal{A}_{0x} = \mathcal{A}_0 \cap T_xM$,
	\item $\mathcal{T}$: a maximally connected conic subbundle where $L > 0$, the $L$-metric exists and has Lorentzian signature $(+,-,-,-)$, with fiber $\mathcal{T}_x = \mathcal{T} \cap T_xM$.
\end{itemize}
The relation between these different subbundles is precisely what distinguishes the different definitions of Finsler spacetimes in the literature mutually among each other. The definition we present here allows for signature changes of the $L$-metric over $TM$, as well as for a measure zero set $TM\setminus \mathcal{A}$ where $g^L$ is non-degenerate or not even defined.

For the application in physics the existence of the subbundle $\mathcal{T}$ and the solvability of the Euler-Lagrange equations are the most important properties. The former condition guarantees the existence of a convex cone of directions in each tangent space $\mathcal{T}_x$ which can be interpreted as timelike directions, while the later one guarantees the existence of freely falling timelike and lightlike trajectories. Wherever the $L$-metric is non-degenerate, i.e.\ on  $\mathcal{A}$, the Euler-Lagrange equations \eqref{eq:EL}, obtained from variation of the Finsler length measure \eqref{eq:Fl} over timelike curves with fixed endpoints, are identical the Finsler geodesic equations
\begin{align}\label{eq:geod}
\ddot x^a + N^a{}_{b}(x,\dot x)\dot x^b = 0
\end{align}
defined in terms of the Cartan non-linear connection coefficients
\begin{align}\label{eq:nlc}
N^a{}_{b}(x,\dot x) = \tfrac{1}{4}\dot{\partial}_b(g^{Lac}(\dot x^d\partial_d\dot{\partial}_c L - \partial_c))\,.
\end{align}
Hence timelike geodesics extremize the Finsler length measure
\begin{align}\label{eq:Fl}
	S[\dot x] = \int \d\tau\ \sqrt{|L|}\,.
\end{align} 

The definition presented here just demands that on Finsler spacetimes there exists a convex cone of timelike directions as well as lightlike directions, along which the Finslerian geometry of spacetime is well-defined. It does not make any statements about the properties of the geometry along other directions. Moreover, it includes Bogoslovsky/Kropina \eqref{eq:bogos}, Randers \eqref{eq:randers} as well as $n$th-root \eqref{eq:nthroot} Lorentzian Finsler geometries. It excludes the Finsler spacetimes constructed in \cite{Caponio:2017lgy,Caponio:2015hca}. The latter have the problem that they do not guarantee that the Finsler metric is non-degenerate along all timelike directions. This in turn means that it is not guaranteed that the curvature tensors, which define the dynamics of a Finsler spacetimes in section \ref{sec:gravi}, are defined along all timelike directions. Hence it is unclear in this setting if a causal evolution of spacetime can be guaranteed for all observers.

The conditions listed here are a minimal set of necessary requirements to be satisfied by a physical Finslerian spacetime geometry. It may turn out in the future that eventually, for physical viability, stronger requirements on Finsler spacetimes are needed. 

Having found a definition of Finsler spacetimes, which ensures the basic notions of causality, demonstrated how to realize one part of the three-fold role of the geometry of spacetime in physics in terms of Finsler geometry.

\section{Observers}\label{sec:obs}
Physical observers are composed out of massive particles and they possess a clock according to the clock postulate. Moreover the are able to measure spatial distances. With these basic operations an observer can perform more general measurements.

\begin{defi}[Observers]\label{def:obs}
	Let $(M,L)$ be a Finsler spacetime and $\gamma:\mathbb{R}\rightarrow R$ be a proper time parametrized curve with timelike tangent $\dot{\gamma}\in \mathcal{T}_\gamma$. Let $P_{(\gamma, \dot{\gamma})} = \frac{1}{2}\dot{\partial}_aL(\gamma, \dot{\gamma}) \d x^a$ be the canonical momentum of $\gamma$. An observer on a Finsler spacetime is a timelike curve $\gamma$, as described, equipped with a unit time direction $e_0$
	and a time-space split, which identifies unit spatial directions $X \in T_\gamma M$, via the conditions:
	\begin{itemize}
		\item $|P_{(\gamma, \dot{\gamma})}(e_0)| = 1$, i.e.\ the observers tangent represents the observers unit time direction $e_0 = \dot{\gamma}$,
		\item co-normality to the observers canonical momentum $1$-form $P_{(\gamma, \dot{\gamma})} = \frac{1}{2}\dot{\partial}_aL(\gamma, \dot{\gamma}) \d x^a$ singles out an observers spatial directions $P_{(\gamma, \dot{\gamma})}(X) = 0$,
		\item the unit spatial directions are normalized operationally by the radar length $\mathcal{L}_{(\gamma, \dot{\gamma})}$ the observer associates to the spatial directions $\mathcal{L}_{(\gamma, \dot{\gamma})}(X) = 1$.
	\end{itemize}
\end{defi}
The identification of the observer's unit time direction $e_0$ with the observer's tangent is a consequence of the clock postulate. The correspondence between the observers spatial directions and the directions which are annihilated by the observers canonical momentum comes from the fact that the integral manifold to these directions locally define the observers equal time surface. The radar length normalization of the spatial directions is chosen since this normalization can operationally be realized by every observer.

The radar length $\mathcal{L}_{(\gamma, \dot{\gamma})}$ an observer on a worldline $\gamma$ associates to a spatial object, is given by the time interval that passes for an observer between emission and reception of a light pulse which propagates from the observer to the end of the object, where it gets reflected, and back. Infinitesimal this experiment can be described by identifying the object with a spatial vector $Z$ and demanding that the vectors $N^\pm = \ell_{(\gamma, \dot{\gamma})}^\pm(Z) \ \dot{\gamma} \pm Z$ are both lightlike, i.e.\ $L(\gamma, N^\pm) = 0$. The radar length the observer associates to $Z$ then is $\mathcal{L}_{(\gamma, \dot{\gamma})}(Z) = \ell_{(\gamma, \dot{\gamma})}^+(Z) + \ell_{(\gamma, \dot{\gamma})}^-(Z)$. It turns out that $\mathcal{L}_{(\gamma, \dot{\gamma})}(Z)$ is a positive definite Finsler function with which the observer measures spatial length. It is connected to the Finsler spacetime Lagrangian $L$ in a highly non-trivial way. Only for $L(x,y) = g_{ab}(x)\dot x^a\dot x^b$ the radar length is simply given by the root of the spacetime Finsler Lagrangian $\mathcal{L}_{(\gamma, \dot{\gamma})}(Z) = \sqrt{|g_{ab}(\gamma)Z^aZ^b|}$. Note that in general $\mathcal{L}_{(\gamma, \dot{\gamma})}(Z) \neq \sqrt{|g^L_{ab}(\gamma,\dot{\gamma})Z^aZ^b|}$. A derivation of the radar length for general Finsler spacetimes can be found in \cite{Pfeifer:2014yua} and its derivation from premetric electrodynamics in \cite{Gurlebeck:2018nme}.

What is missing to speak about an observer frame is a physical relevant relation between three independent spatial observer directions $\{e_\alpha\}_{\alpha=1}^3$. In the usual relativistic observer definition the spatial frame vectors are demanded to be orthogonal with respect to the spacetime metric $g(e_\alpha, e_\beta) = 0$. So far no physical arguments have been presented how the spatial directions of an observer should be related to each other on a Finsler spacetime. The results from studying the radar length normalisation demonstrates that just formally replacing the Lorentzian spacetime metric with the $L$-metric does not yield necessarily a physically operationally meaningful expression, even though it may be mathematically well defined. Thus demanding orthogonality with respect to the $L$-metric may be mathematically well defined, but its physical meaning remains unclear. Hence one of the open questions in the definition of observers on Finsler spacetimes is if there is a good physical operation or principle which identifies a constraint which fixes the spatial frame of an observer.

Other open tasks are the derivation of observables and to find the transformations which map observers, defined as above, onto each other. The latter would lead to the generalisations of local Lorentz transformation of general relativity to Finsler spacetimes.

With this definition of observers, even though it is not yet complete, a first step towards the realisation of the second of the three-fold role of the geometry of spacetime in physics on Finsler spacetime has been achieved. 

\section{Gravitational dynamics}\label{sec:gravi}
There exist basically three strategies to construct Finslerian extensions of the Einstein equations to determine a Finslerian geometry of spacetime dynamically. The first one aims for a tonsorial field equation for the $L$-metric \cite{Asanov,Miron,Chang:2009pa,Voicu:2009wi,Minguzzi:2014fxa}, the second for a scalar field equation for the Finsler Lagrangian \cite{Rutz,Pfeifer:2011xi,Hohmann:2018rpp} and the third aims for field equations for the tensor fields on spacetime from which the Finsler length measure is constructed \cite{Bogoslovsky1994}.

Since the fundamental variable on a Finsler spacetime, from which the geometry is derived, is the Finsler Lagrangian $L$, a scalar on the tangent bundle, the most obvious choice is to look for a scalar field equation, which determines this one scalar field. Moreover, even though this is not necessary but desirable, the field equation shall be the Euler-Lagrange equation of a well-defined action integral. Furthermore, the geometric objects from which the Lagrangian of the action shall be constructed, should be defined by the Finsler Lagrangian alone, without introducing additional fields. Last but not least, the field equation shall reduce to the Einstein equations in the case of a Finsler geometry that is equivalent to a pseudo-Riemannian one, i.e.\ for $L(x,\dot x)=g_{ab}(x)\dot x^a\dot x^b$.

The demand for a scalar field equation singles out the vacuum field equation deduced by Rutz \cite{Rutz} from geodesic deviation on a Finsler spacetime with help of Pirani's argument \cite{Pirani}. From the variational completion algorithm it turned out that Rutz's equation can not be derived from an action. Its variational completion \cite{Hohmann:2018rpp} yields the vacuum field equation which was suggested by the author and collaborator in \cite{Pfeifer:2011xi}
\begin{align}\label{eq:fgrav}
	\frac{1}{2}g^{Lab}\dot{\partial}_a\dot{\partial}_b R - \frac{3}{L} R - g^{Lab}(\nabla_{\delta_a}P_{b} - P_aP_b + \dot{\partial}_a (\dot x^c\nabla_{\delta_c} P_b)) = 0\,.
\end{align}
It can be derived from an action formulated on the positive projective tangent bundle 
\begin{align}
	PTM^+ :=\{[(x,\dot x)]_{\sim}\ |\ (x,\dot x)\in TM\setminus\{0\}\}, (x,\dot x) \sim (x,\dot x')\Leftrightarrow \dot x' = \lambda \dot x\,,
\end{align}
or equivalently, on the unit tangent bundle $\Sigma:= \{(x,\dot x)\in TM\setminus\{0\}\ |\ |L(x,\dot x)|=1\}$. The Lagrangian volume form $\Lambda$ of the action on $\Sigma$ is
\begin{align}\label{eq:fgravL}
	\Lambda = \frac{R}{L} |\det{g^L}|\ \mathbf{i}_{\mathbb{C}}(\d x^0\wedge ... \wedge \d x^3 \wedge \d \dot x^0\wedge ... \wedge \d \dot x^3)\,.
\end{align}
Here $R$ is the canonical non-linear Finsler curvature scalar derived from the non-linear connections coefficients \eqref{eq:nlc} as $R = R^a{}_{ab}\dot x^b = (\delta_a N^a{}_b - \delta_b N^a{}_a)\dot x^b$ and $\delta_a = \partial_a - N^b{}_a\dot{\partial}_b$. The covariant derivative $\nabla_{\delta_a}$ in \eqref{eq:fgrav} is the horizontal Cartan linear covariant derivative defined by the connection coefficients $\Gamma^{CL a}{}_{bc} = \tfrac{1}{2}g^{Laq}(\delta_b g^L_{cq}+\delta_c g^L_{bq}-\delta_b g^L_{cq})$ and the tensor $P_a$ in \eqref{eq:fgrav} denotes the trace of Landsberg tensor $P_a = \dot{\partial}_bN^b{}_a - \Gamma^{CL b}{}_{ba}$. The vector field $\mathbb{C} = \dot x^a\dot{\partial}_a$ is the Liouville vector field and characterizes the normal directions to $\Sigma$. 

For $L(x,y)=g_{ab}(x)\dot x^a\dot x^b$ the action 
\begin{align}
	S = \int_D \Lambda\,,
\end{align}
where $D\subset \Sigma$ compact, is equivalent to the Einstein-Hilbert action and the vacuum field equation becomes equivalent to the vanishing of the Ricci tensor $r_{ab}=0$ of the metric $g$ \cite{Hohmann:2018rpp}.

With this action based Finsler generalisation of the Einstein equations the third of the three-fold role of the geometry of spacetime in physics has been realized. The missing piece to a complete dynamical determination of Finsler spacetimes is the formulation of a consistent physical matter coupling, which is work in progress.

\section{Conclusion and Outlook}\label{sec:concl}

Pseudo-Finsler geometry is capable to serve as physical geometry of spacetime. It provides the three important notions a spacetime geometry has to implement to be the stage where physics takes place: 
\begin{itemize}
	\item Definition \ref{def:fst} guarantees the existence of a causal structure through the identification of causal, timelike and lightlike, directions;
	\item Definition \ref{def:obs} gives a precise notion of observers and their measurements by specifying the time-space split of an observer;
	\item Equation \eqref{eq:fgrav} encodes the gravitational vacuum dynamics, derived from a well defined action principle.
\end{itemize}

An ongoing field of research is to identify the Finsler spacetime geometry which is consistent with all observations from local Laboratory experiments \cite{Itin:2014uia} as well as astrophysics \cite{Perlick:2005hz,Kouretsis:2008ha,Mavromatos:2010nk,Vacaru:2010fi,Li:2012ty,Lammerzahl:2012kw,Basilakos:2013hua,Fuster:2015tua,Hohmann:2016pyt,Lammerzahl:2018lhw}. The most promising observations to detect an imprint of a Finslerian spacetime geometry in astrophysics are the trajectories of light. For example, an energy, frequency or polarisation dependent observation of the time of arrival of gamma-rays emerging from gamma-ray bursts in the early universe, of the shadow of black holes or of gravitational lensing patterns, can be explained by a Finslerian spacetime geometry. In local laboratory experiments Finsler spacetime geometry can be detected by studying physical systems in media. For example the influence of the medium on the propagation of waves and particles or on quantum effects like the Casimir effect, an Unruh detector or quantum energy inequalities can be predicted using Finsler spacetime geometry.

Apart from the systematic comparison between the predictions of a Finsler spacetime geometry and observations, further conceptual questions have to be addressed.

The definition of Finsler spacetimes discussed here is such that it includes the examples of Finsler spacetime geometry encountered in Section \ref{sec:fip} and must be understood as a list of minimal and necessary requirements a pseudo-Finsler geometry has to satisfy at least, in order to be considered as physical geometry of spacetime. Eventually further constraints may be added from observational or conceptual considerations. The most conservative class of Finsler spacetimes to be considered as extension of pseudo-Riemannian geometry are the ones of Berwald type with Lorentzian metric conformal structure \eqref{eq:expg}.

The definition of observers is not yet complete. The important questions to be answered from physical arguments are: How to fix the spatial observer frame? To demand orthogonality with respect to the Finsler metric, evaluated at the observers position on the tangent bundle is one option, which however is more mathematically motivated than physically. And, what are the transformations between observers which generalize local Lorentz transformations to pseudo-Finsler geometry.

The field equations need to be completed by a consistent matter coupling to obtain the source of the Finsler spacetime geometry from the matter fields on spacetime. Here a most promising Ansatz is to consider gases of particle excitations of fields which propagate on a Finsler spacetime geometry. Gases can best be described via the kinetic theory \cite{Ehlers2011}, which is naturally formulated on the tangent bundle in terms of a $1$-particle distribution function. A direct coupling between these fluids and a Finslerian spacetime geometry allows for a finer description of the interaction between the fluid constituents and gravity \cite{Hohmann:2015duq,Hohmann:2015ywa}, compared to the one usually employed via the projection of the fluid dynamics onto spacetime through averaging.

Due to the recent developments in the subject of pseudo-Finslerian spacetime geometry in physics and mathematics discussed in this review, it is exciting to reconsider systematically the application of Finsler spacetime geometry as geometric description of gravity beyond general relativity and as geometric description of physics in media.

\section{Acknowledgments}
Many thanks go to Miguel Sánchez Caja, Miguel Angel Javaloyes, Volker Perlick and Nicoleta Voicu for fruitful discussions and comments which lead to this work. The author acknowledges financial supported by the European Regional Development Fund through the Center of Excellence TK133 ``The Dark Side of the Universe''.



\bibliographystyle{abbrv}
\bibliography{FIP}

\end{document}